\documentclass[aps,prd,amsmath,amssymb,preprintnumbers,onecolumn,11pt,nofootinbib]{revtex4}

\usepackage{mathtools} % mathtools builds on and extends amsmath package
\usepackage[a4paper,left=20mm,right=20mm,top=25mm,bottom=25mm]{geometry}
\usepackage{mathrsfs}
\usepackage{graphicx,epsfig}
\usepackage{color}
\usepackage{fancyhdr}
\usepackage{multirow}
\usepackage{tikz}
\usepackage{feynmp}
\usepackage{mathpazo}
\usepackage{bbold}
\usepackage{natbib}
\usepackage{diagbox} %diagonal in table

\usepackage{amsfonts}
\usepackage{bm}
\usepackage{xcolor}

\definecolor{CP3}{cmyk}{0,0.88,0.77,0.40}
\newcommand\be{\begin{equation}}
\newcommand\ee{\end{equation}}
\newcommand\ba{\begin{eqnarray}}
\newcommand\ea{\end{eqnarray}}

\newcommand\de{{\delta}}
\newcommand{\eff}{\phi}
\newcommand{\ef}{{\rm eff}}

\newcommand\m{{M}}

\newcommand\e{{\rm e}}

\newcommand\al{\alpha}
\newcommand\bt{\beta}

\newcommand\Om{\Omega}
\newcommand\Oms{\Omega_{m\,s}}

\newcommand{\n}{n}

\renewcommand{\(}{\left(}
\renewcommand{\)}{\right)}
\renewcommand{\[}{\left[}
\renewcommand{\]}{\right]}

\newcommand{\lm}{\lambda}

\newcommand{\ga}{\gamma}
\renewcommand{\m}{{\cal M}}

\newcommand\calv{{\cal V}}

\newcommand\alphaB{\alpha_{\text{B}}}

\newcommand\alphaK{\alpha_{\text{K}}}
\newcommand\alphaT{\alpha_{\text{T}}}
\newcommand\alphaL{\alpha_{\text{L}}}
\newcommand\alphaH{\alpha_{\text{H}}}
\newcommand{\quadac}{{(2)}}

\newcommand{\bun}{\beta_1}
\newcommand{\btrois}{\beta_3}
\newcommand{\tg}{\tilde{g}}
\newcommand{\tzeta}{{\tilde{\zeta}}}
\newcommand{\dsz}{\dot{\sigma}_0}

\begin{document}

%%%%%%%%%%%%%%%%%%%%%%%%%%%%%%%%%%%%%%%%%%%%%%%%%%%%%%%%%%%%%%%%%%%%%%%%%%%%%%%%%%%%%%%%

\title{\Large Cosmic evolution in DHOST theory with scaling solutions}
\author{Wittaya Thipaksorn $^{1}$}
\email{wittayat58@nu.ac.th} 
\affiliation{\footnotesize $^{1}${The Institute for Fundamental Study \lq\lq The Tah Poe Academia Institute\rq\rq, \\Naresuan University, Phitsanulok 65000, Thailand}}
\author{Khamphee Karwan$^{1,2}$}
\email{khampheek@nu.ac.th}
\affiliation{\footnotesize $^{1}${The Institute for Fundamental Study \lq\lq The Tah Poe Academia Institute\rq\rq, \\Naresuan University, Phitsanulok 65000, Thailand}}
\affiliation{\footnotesize $^{2}${Thailand Center of Excellence in Physics , Ministry of Higher Education, Science, Research and Innovation, 328 Si Ayutthaya Road, Bangkok 10400, Thailand}}
%%%%%%%%%%%%%%%%%%%%%%%%%%%%%%%%%%%%%%%%%%%%%%%%%%%%%%%%%%%%%%%%%%%%%%%%%%%%%%%%%%%%%%%%

\begin{abstract}

We study cosmic evolution based on the fixed points in the dynamical analysis of the Degenerate Higher-Order Scalar-Tensor (DHOST) theories.
We consider the DHOST theory in which the propagation speed of gravitational waves is equal to the speed of light, the tensor perturbations do not decay to dark energy perturbations, and the scaling solutions exist.
The scaling fixed point associated with late time acceleration of universe can be either stable or saddle depending on the parameters of the theory. 
For some ranges of the parameters,  this scaling fixed point and field dominated fixed point can be simultaneously stable.
Cosmic evolution will reach either the scaling attractor or the field dominated attractor depending on signs of time derivative of the scalar field in the theory during the matter domination.
The density parameter of dark matter can be larger than unity before reaching the scaling attractor if the deviation from the Einstein theory of gravity is too large.
For this DHOST theory, stabilities of $\phi$-matter-dominated epoch ($\phi$MDE) and field dominated solutions are similar to the coupled dark energy models in Einstein gravity even though gravity is described by different theories.
In our consideration, the universe can only evolve from the $\phi$MDE regime to the field dominated regime.
The ghost and gradient instabilities up to linear order in cosmological perturbations have been investigated.
There is no gradient instability, while the ghost instability can be avoided for some range of the parameters of the model.

 %\\[4mm]
\end{abstract}

\maketitle

%%%%%%%%%%%%%%%%%%%%%%
\section{Introduction}

Observed cosmic acceleration \cite{Riess,Perlmutter-2} is one of the important puzzles in modern cosmology which is possible to be explained by supposing that physics of gravity deviates from the Einstein theory on cosmic scales \cite{Clifton:11}.
Deviation from the Einstein theory can be achieved if there are extra degrees of freedom for the gravity in addition to two-tensor degrees of freedom.
For the simplest case, these extra degrees of freedom can be scalar degrees of freedom,
and a class of theories with extra scalar degrees of freedom is scalar-tensor theories of gravity \cite{BransDicke,Horndeski1974,Fujii2003,Galileon2011,Horndeski2012,Horndeski2013,GLPV2014,GLPV2015}.
The DHOST theories which are the most general scalar-tensor theories of gravity are constructed by demanding that the theories are degenerate to eliminate Ostrogradsky instability \cite{DHOST2016:1,DHOST2016:2,DHOST2016:3,DHOST2016:4,DHOST2016:5} .
This class of theories consists of single scalar and two-tensor degrees of freedom for gravity similar to usual Brans-Dicke theory.

The important constraint on The DHOST theories comes from the propagation speed of gravitational waves (GW) which coincides with the speed of light to an accuracy of $10^{-15}$ \cite{Monitor:2017mdv}.
The propagation speed of GW  was measured from The detection of GW and gamma-ray bursts from the merging of a neutron stars binary system \cite{TheLIGOScientific:2017qsa, Coulter:2017wya, GBM:2017lvd,Murguia-Berthier:2017kkn}.
If the propagation speed of GW is required to be always equal to the speed of light,
the action for scalar-tensor theories of gravity is tightly constrained \cite{Lombriser:2015sxa, Lombriser:2016yzn, Bettoni:2016mij,Creminelli:2017sry, Sakstein:2017xjx, Ezquiaga:2017ekz, Baker:2017hug, Arai:2017hxj}.
For the Horndeski action, the non-minimal coupling term that satisfies this constraint is in the form of generalised Brans-Dicke theory.
The action for beyond Horndeski theories \cite{GLPV2015} that satisfies the GW constraint has been discussed and cosmology in this constrained theory has been analyzed in \cite{Kase:2018iwp}.

The cosmic evolution and density perturbation in the DHOST theories which satisfy the constraint on the propagation speed of GW have been studied in various aspects, e.g., \cite{Crisostomi2019:2,Hirano2019:1,Langoi2020:1,Langoi2020:2}.
In addition to the constraint on propagation speed, we demand that GW do not decay to dark energy perturbations \cite{Creminelli2018}.
This requirement together with the constraint on propagation speed of GW tightly constrain form of the Lagrangian for the DHOST theories.
The Vainshtein mechanism for a class of DHOST theories that satisfies these two constraints has been studied in \cite{Hirano2019:2}.

Scaling and tracking behaviours for the cosmic evolution are the interesting features arisen in some models of dark energy and modified theories of gravity,
because they could lead to attractors in the phase space of the cosmic evolution which could satisfy the observational constraints 
\cite{Scaling1998:1,Scaling1998:2,Scaling2000,Scaling2001,Scaling2003:1,Scaling2003:2, DT11b, amen:18}.
A model having scaling behaviour can be obtained by assuming interaction between dark energy and dark matter.
Due to such interaction, a ratio of the energy density of dark energy to that of dark matter is constant with time during the scaling regime. 
A scaling behaviour in the interacting dark energy models has been investigated widely in literature \cite{Scaling2004:1,Scaling2004:2,Scaling2006:1,Scaling2006:2}.
Scaling and tracking solutions in the DHOST theory which satisfy the above two constraints on GW have been discussed.
Demanding existence of the scaling and tracking solutions, the suitable form of the Lagrangians has been derived \cite{Scaling2019}. 

In this work,
we analyze the stabilities of the fixed points found in \cite{Scaling2019},
and discuss cosmic evolution based on these fixed points.
The conditions for avoiding ghost and gradient instabilities in the class of DHOST theories studied in this work are discussed.

In section \ref{sec:2},
we review the DHOST theory that has scaling solutions.
The fixed points of the cosmic evolution and their stabilities are analyzed in section \ref{sec:3}.
The possibles cosmic evolution  associated with these fixed points are discussed .
In section (\ref{sec:4}), we study stability of the linear cosmological perturbations around cosmological background.
We conclude in section (\ref{sec:5}).

\section{DHOST theories with scaling solutions}
\label{sec:2}

\subsection{Review on DHOST theories}

The DHOST theories are constructed by imposing the degeneracy conditions to the most general form of Lagrangian containing second order derivatives of scalar field in the form
\be
L = G_2(\phi, X) + G_3(\phi, X) \Box\phi + G_4(\phi, X) R 
+ C_{(2)}^{\al\bt\mu\nu}\phi_{\al\bt}\phi_{\mu\nu} 
+ C_{(3)}^{\al\bt\mu\nu\rho\sigma}\phi_{\al\bt}\phi_{\mu\nu}\phi_{\rho\sigma}\,,
\label{lang1}
\ee
where $R$ is the Ricci scalar, $X \equiv - \phi_\mu\phi^\mu$,
$\phi_\mu \equiv \nabla_\mu\phi$, $\phi_{\mu\nu} \equiv \nabla_\nu\nabla_\mu \phi$,
and $\nabla_\nu$ denotes covariant derivative compatible with the metric $g_{\mu\nu}$.
In the following consideration, we concentrate on the terms up to the quadratic in the second order derivatives of scalar field.
The possible form of the quadratic terms can be written as \cite{DHOST2016:1}
\ba
C_{(2)}^{\al\bt\mu\nu}\phi_{\al\bt}\phi_{\mu\nu}  
&=&
A_1(\phi, X) \phi_{\mu\nu}\phi^{\mu\nu} + A_2(\phi, X)(\Box\phi)^2 +A_3(\phi, X)\Box\phi \, \phi^\mu\phi_{\mu\nu}\phi^\nu\
  \nonumber\\
  && 
+ A_4(\phi, X)\phi^\mu \phi_{\mu\rho}\phi^{\rho\nu}\phi_\nu\ + A_5(\phi, X)(\phi^\mu\phi_{\mu\nu}\phi^\nu)^2\,.
\label{c2}
\ea
Based on the degeneracy conditions, the DHOST theories can be classified into three classes.
However, the theories in the class I can be free from the gradient instability while the class II cannot because the sound speed square of the tensor and scalar perturbations have opposite sign \cite{Langoi2017,Langoi2018}
For the class III, the tensor degrees of freedom do not propagate.
Hence, we concentrate on the class I.
The degeneracy conditions for the class I are
\ba
A_2 &=& - A_1\,,
\label{deg:a2}\\
A_4 &=& -A_3 + \frac{\left(-4G_{4X} - 2A_1 - XA_3\right)\left(-12G_4G_{4X} - 6A_1 G_4 - 8A_1^2X + A_3G_4 X - 16A_1G_{4X}X\right)}{8(G_4 + XA_1)^2}\,,
\label{deg:a4}\\
A_5 &=& \frac{\left(-4G_{4X} - 2A_1 - XA_3\right)\left(-2A_1^2+3XA_1A_3-4G_{4X}A_1+4G_4 A_3\right)}{8(G_4+XA_1)^2}\,,
\label{deg:a5}
\ea
where subscript ${}_X$ denotes derivative with respect to $X$.
For the DHOST theories which depend quadratically on second order derivatives of scalar field,
the propagation speed of the tensor perturbations is given by \cite{deRham2016}
\be
c_T^2 = \frac{G_4}{G_4 + XA_1}\,,
\label{ct2}
\ee
where the speed of light is equal to unity in this expression.
From the LIGO/VIRGO results \cite{Monitor:2017mdv,Scaling2019,Crisostomi2018},
$c_T$ is equal to speed of light, 
so that  Eq.~(\ref{ct2}) yields
\be
A_1 =0\,.
\label{gw:a1}
\ee
It has been shown that the GW in DHOST theories can decay to scalar perturbations.
To avoid such decay, we demand \cite{Creminelli2018}
\be
A_3 =0\,.
\label{gw:a3}
\ee
Inserting the conditions from Eqs.~(\ref{gw:a1}) and (\ref{gw:a3}) into Eqs.~(\ref{deg:a4}) and (\ref{deg:a5}) , we get 
\be
A_5 = 0\,,
\quad\mbox{and}\quad
A_4 = \frac{6 G_{4X}^2}{G_4}\,.
\label{a4fin}
\ee
Hence, the action for quadratic DHOST theories in which the propagation speed of GW is equal to speed of light and the GW do not decay to dark energy perturbations can be written in the form
\be
S_G = \int d^4x \sqrt{-g} \left\{
G_2 + G_3\Box\phi
+ G_4 R 
+ \frac{6 G_{4X}^2}{G_4} \phi^{\mu} \phi_{\mu \rho} \phi^{\rho \nu} \phi_{\nu}
\right\}\,,
\label{actGW2}
\ee
where we have set the reduced Planck mass $M_p \equiv 1/\sqrt{8\pi G} = 1$.
The $G_3$ term will be dropped in the following consideration for simplicity.
The above action can also be obtained by applying a conformal transformation in which transformation coefficient depends on both a scalar field and its kinetic term to the Einstein-Hilbert action \cite{Zumalaca}.
We write the total action as 
\be
S = S_G + S_\m\,.
\label{actGW2-sm}
\ee
Here,  $S_\m$ is the action for the matter in the universe including ordinary matter and dark matter.
The ordinary matter is minimally coupled to gravity,
while the dark matter  is coupled non-gravitationally to scalar field $\phi$ and coupled non-minimally to  gravity. 
The action for dark matter is denoted by $S_m$ in the subsequent sections.

\subsection{Evolution equations for the background universe}

To study the evolution of the background universe in the DHOST theories described by the action (\ref{actGW2-sm}),
we use the Friedmann-Lema\^{i}tre-Robertson-Walker (FLRW) metric for the spatially flat universe in the form
\be
ds^2 = - n^2(t)dt^2 + a^2(t)\delta_{ij}dx^i dx^j\,,
\ee
where $\delta_{ij}$ is the Kronecker delta,
$a(t)$ is the cosmic scale factor and
$n(t)$ is an auxiliary function which will be set to unity after the evolution equations are obtained.
Using the above line element and homogeneity of the scalar field in the background universe,
the action (\ref{actGW2-sm}) becomes
%%%%
\be
 S = \int dt a^3 n\left\{
G_2- 6 G_{4\phi} H \frac{\dot\phi}{n^2}
-6 G_4\[\frac{H}{n} + \frac{G_{4X}}{G_4} \frac{\dot\phi}{n^2}\frac{d}{dt}\(\frac{\dot\phi}{n}\) \]^2
\right\}
+ S_\m\,,
\label{action-homo}
\ee
%%%%
where a dot denotes derivative with respect to time $t$, $H \equiv \dot a / a$ is the Hubble parameter,
and subscript ${}_\phi$ denotes derivative with respect to $\phi$.

Variations of the action (\ref{action-homo}) with respect to $n$ and $a$ yield
%%%%
\ba
\rho_\m &=& E_{00} \equiv   \frac{1}{G_4^2}\left[-G_4 X \left(-6 \dot{\phi } \left(-2G_{4 X}^2
    \dddot\phi - 6 H G_{4 X}^2 \ddot{\phi}\right)+G_4 \left(12 \left(2 H^2+\dot{H}\right) G_{4 X}+2 G_{2X}\right) + 6 G_{4 X}^2 \ddot{\phi}^2\right)
    \right. 
   \nonumber\\
   && \left. 
    +G_4^2 \left(6 G_4 H^2+6 H \dot{\phi } \left(2 G_{4 X} \ddot{\phi }+G_{4 \phi }\right)+G_2\right)+12 X^2 G_{4 X}
    \ddot{\phi } \left(\left(G_{4 X}^2-2 G_4 G_{4 X X}\right)
    \ddot{\phi }-2 G_4 G_{4 \phi X}
   \right. \right. 
    \nonumber\\
   && \left.\left. 
    +G_{4 X} G_{4 \phi}\right)\right] \,,
\label{e00}
\ea
and
\ba
- p_\m &=& E_{ii} \equiv  \frac{1}{G_4} \left[G_4 \left(4 \dot{\phi } \left(G_{4 X}\dddot\phi + 2 H G_{4 X} \ddot{\phi }
   + H G_{4 \phi }\right)+6 G_4 H^2+4 G_4 \dot{H}+4 G_{4
    X} \ddot{\phi }^2+2 G_{4 \phi } \ddot{\phi }+G_2\right)
      \right.
\nonumber\\
&&   \left.
  +X \left(\left(8 G_4 G_{4 X X}-6 G_{4 X}^2\right) \ddot{\phi }^2+8 G_4
    \ddot{\phi } G_{4 \phi  X}+2 G_4 G_{4 \phi  \phi }\right)\right] \,,
\label{eii}   
\ea
%%%%
where $\rho_\m $ and $p_\m$ are the energy density and pressure of the total matter which each of the matter components is perfect fluid.
These quantities can be obtained by varying the action $S_\m$ with respect to metric.
Eqs.~(\ref{e00}) and (\ref{eii}) together with Eq.~(\ref{rpeff}) agree with Eqs.~(2.13) and (2.14) in \cite{Scaling2019}.

 These two equations can be combined to eliminate $\dot{H}$ as
%%%%
\ba
0&=&    \frac{1}{G_4^2} \left[G_4 X \left(-6 G_4 H^2 G_{4 X}+6 H \dot{\phi } \left(2 G_{4 X} G_{4
    \phi }-2 G_{4 X}^2 \ddot{\phi }\right)+6 G_{4 X}^2 \ddot{\phi }^2+6 G_{4 X} G_{4 \phi } \ddot{\phi }-2 G_4 G_{2
    X}+3 G_2 G_{4 X}\right)
\right.
\nonumber\\
&& \left.  
    + G_4^2 \left(6 G_4 H^2+6 H \dot{\phi} \left(2 G_{4 X} \ddot{\phi }+G_{4 \phi }\right)+G_2\right)-G_4 \rho_\m
    \left(G_4-3 X G_{4 X} w_\m\right)
\right.
\nonumber\\
&& \left.    
    +3 X^2 G_{4 X} \left(-2 G_{4 X}^2 \ddot{\phi }^2+4 G_{4 X} G_{4 \phi } \ddot{\phi }+2 G_4 G_{4
    \phi  \phi }\right)\right]\,.
\label{e0:ori}
\ea
%%%%
In the above equation, $w_\m \equiv p_\m / \rho_\m$ is the equation of state parameter of the total matter which is not necessarily zero. 
Varying the action (\ref{action-homo}) with respect to scalar field $\phi$,
we get the evolution equation for scalar field which can be written in the form
\be
F(\ddddot\phi, \dddot\phi, \ddot\phi,\dot\phi,\phi, \ddot{H}, \dot{H}, H)
 = Q\,,
\label{kg}
\ee
where $Q$ is the interaction term arisen from the variation of the action $S_m$ for dark matter  with respect to scalar field $\phi$.
We will see in the following sections that The coupling between the scalar field and matter is needed for shifting the effective equation of state parameter $w_\ef \equiv w_\eff \Omega_\eff$ during the scaling regime at late time to a negative value as required by observations.
Here, $w_\eff$ and $\Omega_\eff$ are the effective equation of state parameter and effective density parameter of scalar degree of freedom associated with scalar field $\phi$ defined below.
In the following consideration,
we suppose that the interaction term $Q$ is a consequence of a energy transfer between scalar field and dark matter.
In principle, the form of the interaction term $Q$ depends on the form of $S_m$.
However, for simplicity, we start with the phenomenological form of the interaction term studied in the literature.
We write the function $F$ in the above equation in the form of the conservation equation for the effective energy density of the scalar field as $F \to \dot\rho_\eff + 3 H (\rho_\eff + p_\eff) = 0$.
Then we add the phenomenological interaction term  on the right-hand side of the conservation equation as
%%%%
\be
\dot\rho_\eff + 3 H(\rho_\eff + p_\eff) = - Q \rho_m \dot\phi\,,
\label{drhop}
\ee
%%%%
where $Q$ is constant, $\rho_\eff$ and $p_\eff$ are the effective energy density and the effective pressure of the scalar field $\phi$.
Supposing that the total  energy density of the scalar field and dark matter is conserved,
we have
\be
\dot\rho_m + 3 H \rho_m =  Q \rho_m \dot\phi\,,
\label{rmdot}
\ee
where a subscript ${}_m$ denotes the quantities for dark matter.
The effective energy density and pressure of the scalar field are defined such that Eqs.~(\ref{e00}) and (\ref{eii}) take the forms of the usual Friedmann and acceleration equations as $3 H^2 = \rho_\m + \rho_\eff$ and $2\dot H + 3 H^2 = -p_\m - p_\eff$.
The expressions for $\rho_\eff$ and $p_\eff$ can be read from Eqs.~(\ref{e00}) and (\ref{eii}) as
\be
\rho_\eff \equiv  3 H^2 - E_{00}\,,
\quad 
p_\eff \equiv E_{ii} - 2 \dot{H} - 3 H^2\,.
\label{rpeff}
\ee
From the above expressions, the effective equation of state parameter of the scalar field can be defined as $w_\eff \equiv p_\eff / \rho_\eff$.
According to the definitions of $p_\eff$ and $ \rho_\eff$,
we can write
\be
\frac{\dot H}{H^2} = -\frac 32 \(1 + \Omega_\m w_\m + \Om_\eff w_\eff \) = -\frac 32 \(1 + \frac 13 \Omega_\gamma +  w_\ef \) \,.
\label{weff}
\ee
Here, $\Omega_\gamma$ is the density parameter of radiation where $\Omega_\m = \Omega_m + \Omega_\gamma$ and $\Omega_m$ is the density parameter of dark matter.
Evolution of the background universe can be studied using the dynamical analysis.
% In the dynamical analysis, energy density  of baryon is ignored.
To compute the autonomous equations describing evolution of the background universe,
we have to know the expression for the ratio $\dot H / H^2$.
To compute this ratio, we firstly differentiate Eq.~(\ref{eii}) with respect to time.
Then we eliminate $\ddddot\phi$ from the resulting equation using Eq.~(\ref{kg}).
The remaining $\dddot\phi$ terms can be eliminated using eq.~(\ref{eii}).
Finally, we obtain
\be
0 = \tilde{E}_i(\ddot\phi, \dot\phi,\phi, H, \rho_m)\,.
\label{eti}
\ee
Differentiating the above equation with respect to time and eliminating $\dddot\phi$ terms using Eq.~(\ref{eii}),
we get
\be
\frac{\dot H}{H^2} = - h(\ddot\phi, \dot\phi,\phi, H, \rho_m)\,.
\label{dhst}
\ee

\subsection{Scaling solutions}

The gravity theories described by the action (\ref{action-homo}) can have the scaling behaviour if the effective energy density and the effective pressure of the scalar field together with the energy density of the matter obey the scaling relations $\rho_\eff \propto p_\eff \propto \rho_m \propto H^2$.
During the scaling regime,
we have
\be
\frac{\dot\phi}{H} = \frac{2 h}{\lambda} ={\rm constant} \,.
\label{scalingSol}
\ee
Based on the analysis in \cite{Scaling2019}, for constant coupling $Q$,
the DHOST theory in action (\ref{action-homo}) has scaling solutions if
%%%%
\be
G_2 =  X g_2\left(Y\right) \qquad G_4 = \frac{1}{2}+ g_4 \left(Y\right) \,,
\label{g2g4:scaling}
\ee
%%%%
where $G_2$ and $G_4$ are arbitrary functions of
\be
Y = \frac{X \e^{\frac{\lambda \phi}{M_p}}}{M_p^4} \,.
\ee
Here, $\lambda$ is constant and $M_p$ is restored in the above expression and some of subsequent relations to avoid confusion.
To study scaling solutions in DHOST theories ,
we set $G_2$ and $G_4$ according to Eq.~{g2g4:scaling}.
Since $Y$ is a linear function of the kinetic term $X$,
we choose $G_2$ and $G_4$ as polynomial functions of $Y$ \cite{Scaling2019}:
\ba
\label{G2}
G_2 &=& X \left(\tilde{c}_2  Y^{n_2} - \tilde{c}_6 Y^{n_6}\right)\,, \\
G_4 &=& \frac{1}{2} + \tilde{c}_4 Y^{n_4} \,,
\label{G4}
\ea
where $\tilde{c}_2$, $\tilde{c}_4$ and  $\tilde{c}_6$ are constant and $n_2$, $n_4$ and $n_6$ are constant integers.
The above expressions for  $G_2$ and $G_4$ are mainly motivated by demanding scaling solutions in the model.
Additional motivation for these choices of $G_2$ and $G_4$ are based on the requirement that $G_2$ can reduce to a form of the Lagrangian for canonical scalar field if $n_2 = 0$ and $n_6 = -1$,
while $G_4$ can reduce to case of the Einstein theory when $\tilde{c}_4 = 0$.
Since the existence of the scaling solutions requires that $G_2$ and $G_4$ depend on the scalar field $\phi$ through $Y$,
shift symmetry is broken in this scaling model and consequently a self-accelerating solution is absent.
This implies that the field $\phi$ is required to slowly varying with time to drive the cosmic acceleration at late time.
When the coupling between scalar field and dark matter is constant, the scaling solutions can give
\be
\lambda = -\frac{2 h Q}{3 \Omega_\eff w_\eff} \,.
\label{scalingSol2}
\ee
%%%%

\section{Stabilities of the fixed points and the corresponding cosmic evolution}
\label{sec:3}

\subsection{Autonomous equations}

To compute the autonomous equations from the evolution equations in the previous section, we define the dimensionless variables as 
\be
x \equiv \frac{\dot{\phi}}{M_p H}\,,
\quad
y \equiv  \frac{M_p^2 \e^{\frac{- \lambda \phi}{M_p}}}{H^2} \,,
\quad
z \equiv  \frac{\ddot{\phi}}{\dot{\phi} H}\,.
\label{dless}
\ee
For convenience, we normalize the variables $x$, $y$ and $z$ by their values at scaling fixed point, such that
\be
x_r \equiv \frac{x}{x_s}\,,
\quad
y_r \equiv \frac{y}{y_s}\,,
\quad
\mbox{and}
\quad
z_r \equiv \frac{z}{z_s}\,,
\ee
where subscript ${}_s$ denotes the quantities at the scaling fixed point,
The scaling fixed point in this case is the fixed point that $x$ satisfies the condition in Eq.~(\ref{scalingSol}) and $Q$ satisfies Eq.~(\ref{scalingSol2}).
To compute $x_s$ and $z_s$, we compute derivative of $x$ with respect to $N \equiv \ln a$
as 
\be
x' = z x - x \frac{\dot H}{H^2} \,,
\label{xp}
\ee
which is a possible form of the autonomous equation.
Here, a prime denotes derivative with respect to $N$.
From the condition in Eq.~(\ref{scalingSol}), we have
\be
h_s = \left. \frac{\dot \phi \lm}{2 H}\right|_s \equiv \frac{x_\lm}{2}  \,,
\label{dh2h2s}
\ee
where $x_\lm  \equiv x_s \lm$.
Inserting this solution into Eq.~(\ref{xp}), we get $z_s = - h_s = -x_\lm / 2$.
In terms of dimensionless variables, the constraint equations (\ref{e0:ori}) and (\ref{eti}) are given by Eqs.~(\ref{e0less}) and (\ref{e1less}) in the appendix.
We see that these constraint equations can be solved for $z$ and $\Omega_m$ in terms of $x$ and $y$.
Here we are interested in the evolution of the late-time universe so that we set $\Omega_\gamma = 0$. 
Hence, the late-time dynamics of the background universe can be described  by two dynamical variables $x$and $y$.

Using definitions of $x_r$ and $y_r$,
we can write the autonomous equations as
\ba
x_r' &=& - \frac{x_{\lambda} z_r x_r}{2} - x_r \frac{\dot{H}}{H^2}\,,
\label{xrp}\\
y_r' &=& - x_{\lambda} x_r y_r - 2 y_r \frac{\dot{H}}{H^2} \,,
\label{yrp}
\ea
where $z_r$ is computed from the constraint equations which the solutions are shown in Eqs.~(\ref{xr4sol1})--(\ref{xr4sol3}).
When the autonomous equations are written in these forms,
the coupling constant $Q$ in the autonomous equations is always divided by $\lambda$ so that dynamics of the background universe depend on $Q/\lambda$ rather than $Q$.
In the numerical integration for the evolution of the universe discussed below,
we concentrate on the case where $z_r$ is the first solution given in Eq.~(\ref{xr4sol1}) to avoid contributions from  imaginary part of the solution.
We note that the solutions which give $z_r = x_r = y_r = 1$ are not necessarily be the solution in Eq.~(\ref{xr4sol1}) unless $n_4 = \pm 1$. 
Hence, in our numerical integration for the cosmic evolution, we set $n_4$ to be either $-1$ or $1$.
According to Eq.~(\ref{dhst}), $\dot H / H^2$ also depends on $\Omega_m$.
However $\Omega_m$ in this expression can be eliminated using the constraint equations Eq.~(\ref{e0less}).

To compute the fixed points of this system, 
we set $x_r$, $y_r$ and $z_r$ in the constraint equations Eqs.~(\ref{eiiless}) and (\ref{e0less}) to be unity,
and then we solve for the parameters as
\ba
 c_2 &=&  -\frac{1}{2 \left(2 c_4+1\right)^2\left(n_2-n_6\right)}\left[-6 c_4^2 \left(-2 \left(\Oms + 2 n_6 \left(x_{\lambda}
	  -3\right)+x_{\lambda }-6\right)+2 n_4^3 x_{\lambda }^2
    \right.\right.
    \nonumber\\
    && \left.\left.  
	  -n_4^2 x_{\lambda }\left(n_6 x_{\lambda }+x_{\lambda }-6\right)+4 n_4 \left(x_{\lambda}-4\right)\right)
	  +6 c_4 \left(2 \Oms - n_4 \left(x_{\lambda}-4\right)+2 n_6 \left(x_{\lambda }-3\right)+x_{\lambda }-6\right)
    \right.
    \nonumber\\
    && \left. 
	  -4 c_4^3\left(3 n_4^3 x_{\lambda }^2-3 n_4^2 x_{\lambda } \left(n_6 x_{\lambda}+x_{\lambda }-6\right)
	  +6 n_4 \left(x_{\lambda }-4\right)-2 \left(2 n_6\left(x_{\lambda }-3\right)+x_{\lambda }-6\right)\right)+3 \Oms
    \right.
    \nonumber\\
    && \left.   
	  +2 n_6 \left(x_{\lambda }-3\right)+x_{\lambda }-6\right] \, ,
 \label{Eqc2}
\ea
\ba
c_6 &=&  -\frac{1}{2 \left(2 c_4+1\right)^2 \left(n_2-n_6\right)} \left[6 c_4^2 \left(2 \left(\Oms+x_{\lambda }
	 -6\right)-2 n_4^3x_{\lambda }^2+n_4^2 \left(x_{\lambda }-6\right) x_{\lambda }-4 n_4 \left(x_{\lambda }-4\right)
    \right.\right.
    \nonumber\\
    && \left.\left.
	 +n_2 \left(n_4^2 x_{\lambda }^2+4 x_{\lambda}-12\right)\right)+6 c_4 \left(2 \Oms - n_4 \left(x_{\lambda}
	 -4\right)+2 n_2 \left(x_{\lambda }-3\right)+x_{\lambda }-6\right)
    \right.
    \nonumber\\
    && \left. 
	 -4 c_4^3\left(3 n_4^3 x_{\lambda }^2-3 n_4^2 \left(x_{\lambda }
	 -6\right)x_{\lambda }+6 n_4 \left(x_{\lambda }-4\right)+n_2 \left(-3 n_4^2 x_{\lambda }^2-4 x_{\lambda }+12\right)-2 \left(x_{\lambda}
	 -6\right)\right)
    \right.
    \nonumber\\
    && \left.
	 +3 \Oms + 2 n_2 \left(x_{\lambda }-3\right)+x_{\lambda}-6\right] \,,
 \label{Eqc6}
\ea
where $\Oms$ is $\Omega_m$ at the scaling fixed point,
and we redefine the coefficients as 
\be
c_2 \equiv \tilde{c}_2 x_s^2 Y_s^{n_2}\,, 
\quad 
c_4 \equiv \tilde{c}_4 Y_s^{n_4}\,, 
\quad 
\mbox{and}
\quad
c_6 \equiv \tilde{c}_6 x_s^2 Y_s^{n_6} \,.
\ee
%%%%%
We set $h_s = x_\lm / 2$ and $x_r = y_r = 1$,
and substitute $c_2$ and $c_6$ from Eq.~(\ref{Eqc2}) and (\ref{Eqc6}) into Eq.~(\ref{dhst}) as 
\be
  \frac{x_\lm}{2} = h(\ddot\phi, \dot\phi,\phi, H, \rho_m)|_s = h(x_r, y_r, z_r, \Omega_m)|_s = h(1,1,1, \Oms)\,.
\ee
This relation yields 
\be
0=\frac{18 c_4 \left(2 c_4+1\right){}^4 n_4 \Oms (Q_\lambda-2)
    x_{\lambda }^{13} \left(Q_\lambda x_{\lambda }+x_{\lambda
    }-3\right)}{\lambda ^{12}} \, ,
\label{fix:dh}
\ee
where $Q_\lambda = Q/\lambda$.
The interesting conditions required by the above equation are
\be
\Oms = 0\,, \quad
Q_\lambda x_{\lambda }+x_{\lambda}-3 = 0 \,,
\quad
\mbox{or}
\quad
c_4 = 0 \,.
\label{conditions}
\ee
We can see that $Q_\lambda-2 = 0$ is the special case of the condition $Q_\lambda x_{\lambda }+x_{\lambda}-3 = 0$ .
These conditions lead to three classes of fixed point as follows :
(1) $Q_\lambda x_{\lambda }+x_{\lambda}-3 = 0$ corresponding to scaling fixed point where $Q$ satisfies Eq.~(\ref{scalingSol2}),
(2) $\Oms = 0$ corresponding to the field dominated point where $Q$ does not necessarily satisfy Eq.~(\ref{scalingSol2}),
and (3) $c_4 =0 $ yielding $y_r=0$ for negative $n_4$. 
These fixed points have been found in \cite{Scaling2019}. 
The stabilities of these fixed points will be discussed in the next section.

\subsection{Fixed points and stabilities}

To investigate stabilities of the fixed points,
we linearize the autonomous equations around the fixed point and check signs of the eigenvalues of the Jacobian matrix defined by
\be
J_{ij} = \left. \frac{\partial x_i'}{\partial x_j}\right|_{\rm fixed \, point}\,,
\ee
where $x_i = (x_r,y_r)$.

\subsubsection*{(a) Scaling fixed point}

The scaling fixed point corresponds to the condition 
\be
x_\lm = \frac{3}{Q_\lm + 1} \,.
\label{scalingcon}
\ee
From $h_s = x_\lm/2$, we have
\be
w_\ef = -\frac{Q_\lm}{Q_\lm + 1} \,.
\label{wef:sc}
\ee
We see that if the coupling term disappears, $w_\ef = 0$ because for the scaling solution $\rho_\phi / \rho_m $ is constant.
Using the relation $w_\ef = \Omega_\eff w_\eff$ and Eq.~(\ref{wef:sc}),
we can compute $\Omega_\eff$ as well as $\Omega_m$ at the fixed point if $w_\eff$ at the fixed point is specified.
Inserting the relations for the scaling fixed point into the Jacobian matrix,
we obtain the polynomial equation for the eigenvalues of the fixed point. 
For the sufficiently large $c_4$, the eigenvalues of the Jacobian matrix depend only on  $x_\lambda$ and  given by
%%%%
\be
E_{a\,l} = \left\{\frac{x_\lambda - 6}{2}, 0 \right\}\,.
\label{ea-lc4}
\ee
Since one of the eigenvalues is zero,
the stability of this fixed point cannot be determined using the linear stability analysis.
Non-linear stability analysis can be performed using the center manifold method,
but we will not consider  the non-linear analysis in this work.
If $c_4$ is not too large, the eigenvalues of the Jacobian matrix can be written as
%%%%
\be
E_a = \left\{\mu_1, \mu_2\right\}\,.
\label{ea}
\ee
To describe the accelerated expansion of the late-time universe required by observations, we demand $x_\lambda < 1$.
The eigenvalues $\mu_1$ and $\mu_2$ can be computed from the equation
\be
a_2 \mu^2 + a_1 \mu + a_0 = 0\,,
\ee
where $a_2, a_1$ and $a_0$ are complicated functions of $x_\lambda, \Omega_{m\,s}, c_2, c_4, c_6, n_2, n_4$ and $n_6$.
The solutions for the above equation can be written as
\be
\mu_1 = \frac{x_\lambda- 6}{4} \(1 - \sqrt{1 + \frac{8 a_0}{a_1 (x_\lambda - 6)}}\)\,,
\quad
\mu_2 = \frac{x_\lambda- 6}{4} \(1 + \sqrt{1 + \frac{8 a_0}{a_1 (x_\lambda - 6)}}\)\,.
\label{eigen}
\ee
In the above expressions, the relation $a_1 / (2 a_2) = (6 - x_\lambda)/4$ is used.
It follows from the relations for $\mu_1$ and $\mu_2$ that the real part of $\mu_2$ is always negative for $x_\lambda <6$,
while real part of $\mu_1$ can be either negative or positive.
Hence, the fixed point is stable when the real part of $\mu_1$ is negative and becomes saddle when the real part of $\mu_1$ is positive. 
Due to the lengthy expressions for $a_0, a_1$ and $a_2$, we compute $\mu_1$ numerically and plot the results as a function of $c_4$.

\begin{figure}
\includegraphics[height=0.39\textwidth, width=0.49\textwidth,angle=0]{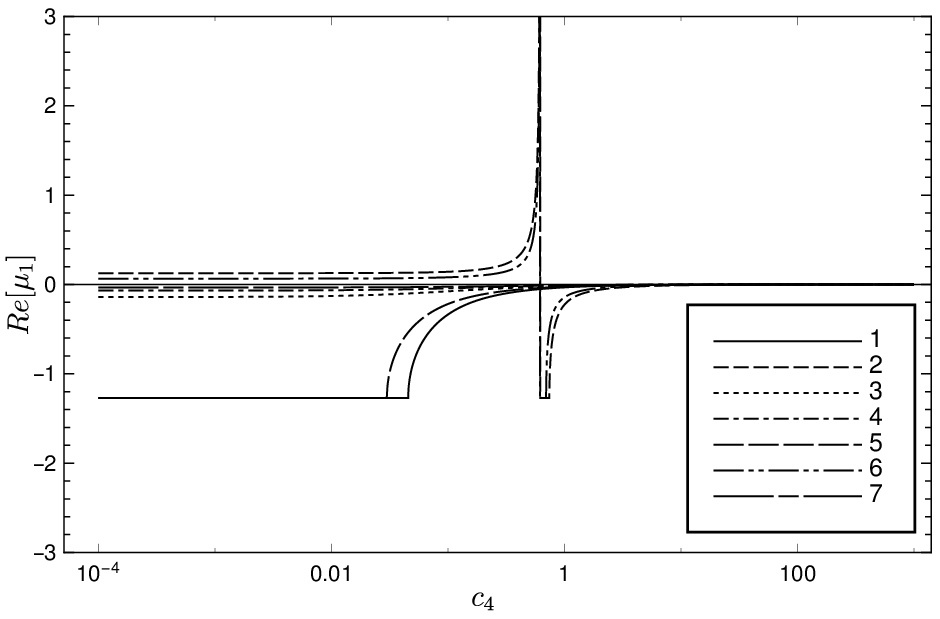}
\includegraphics[height=0.39\textwidth, width=0.49\textwidth,angle=0]{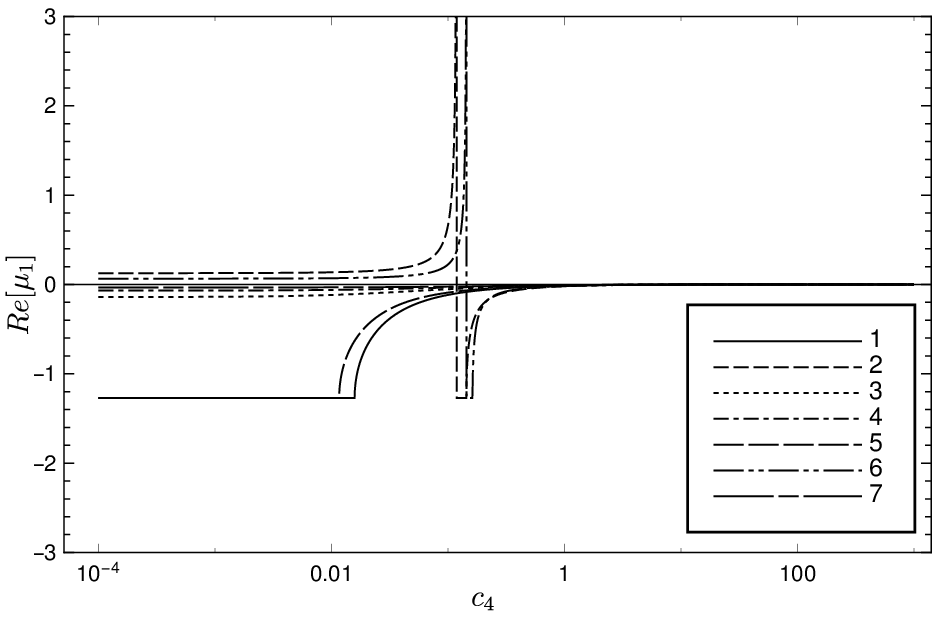}\\
\includegraphics[height=0.39\textwidth, width=0.49\textwidth,angle=0]{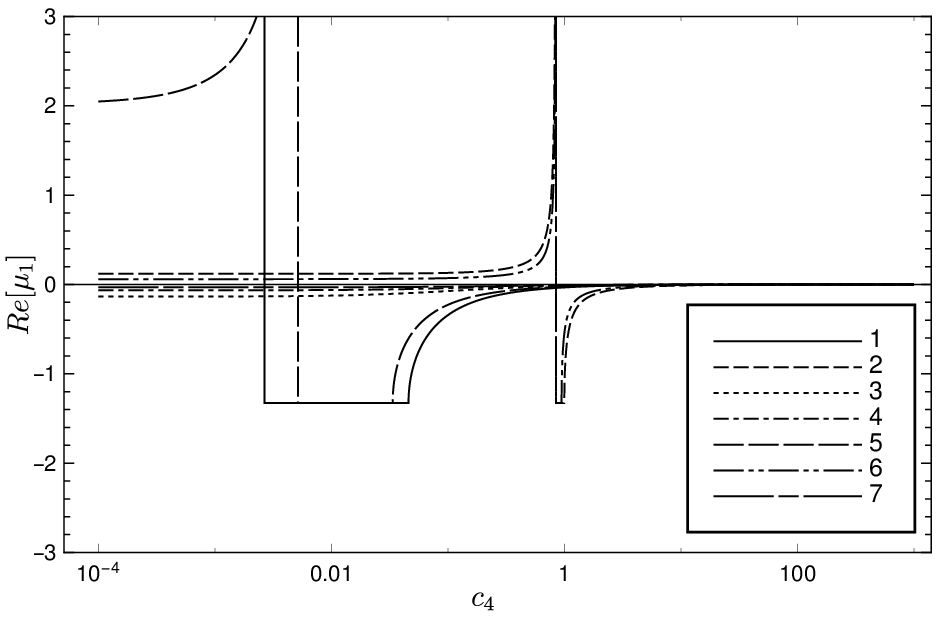}
\includegraphics[height=0.39\textwidth, width=0.49\textwidth,angle=0]{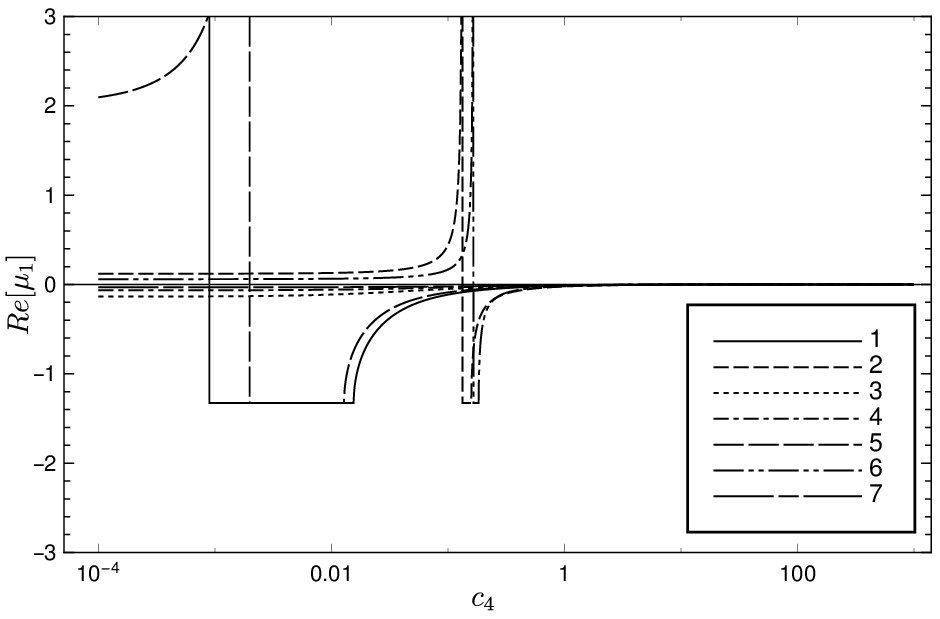}
\caption{\label{fig:1}
Plots of the real part of $\mu_1$ as a function of $c_4$. The upper left, upper right, lower left and lower right panels correspond to 
$(x_\lambda, n_4) = (0.92,-1), (0.92,-2), (0.69,-1)$ and $(0.69,-2)$, respectively. In the plots, lines 1, 2 ,3 , 4, 5, 6 and 7 represent the cases 
of $(n2,n6)$ = (0,-1), (0,-3), (0,1), (0,3), (1,-1), (1,-3) and (1,3). 
}
\end{figure}

The real part of $\mu_1$ for some choices of the parameters is plotted in Fig.~(\ref{fig:1}).
In all plots, $x_\lambda $ and $\Omega_{m\,s}$ are chosen such that $w_\ef$ satisfies observational constraints.
For $\Omega_{m\,s} = 0.3$, we set $x_\lambda = 0.92$ and $x_\lambda = 0.69$ which correspond to $w_\eff = -0.99$ and  $w_\eff = -1.10$, respectively.
From Fig.~(\ref{fig:1}) and Eq.~(\ref{eigen}), we see that the stability of the fixed point depends on $x_\lambda$ which controls the value of $w_\ef$ through the relation $x_\lambda = -3 (1+ w_\ef)$.
In the plot, when $x_\lambda$ decreases, the fixed point of some models, e.g., the model with $ n_6 = -1$, can become saddle point.
According to Fig.~(\ref{fig:1}), the fixed point is stable for the wide range of $c_4$ if $n_6$ is positive.
For $n_6 = -3$, the fixed point can be either saddle or stable depending on the value of $c_4$.
From the plot, we see that the real part of $\mu_1$ reaches zero when $c_4$ is sufficiently large independent of $n_2, n_4, n_6$ and $x_\lambda$,
which agrees with Eq.~(\ref{ea-lc4}).

\subsubsection*{(b) Field dominated point}

\begin{figure}
\includegraphics[height=0.39\textwidth, width=0.49\textwidth,angle=0]{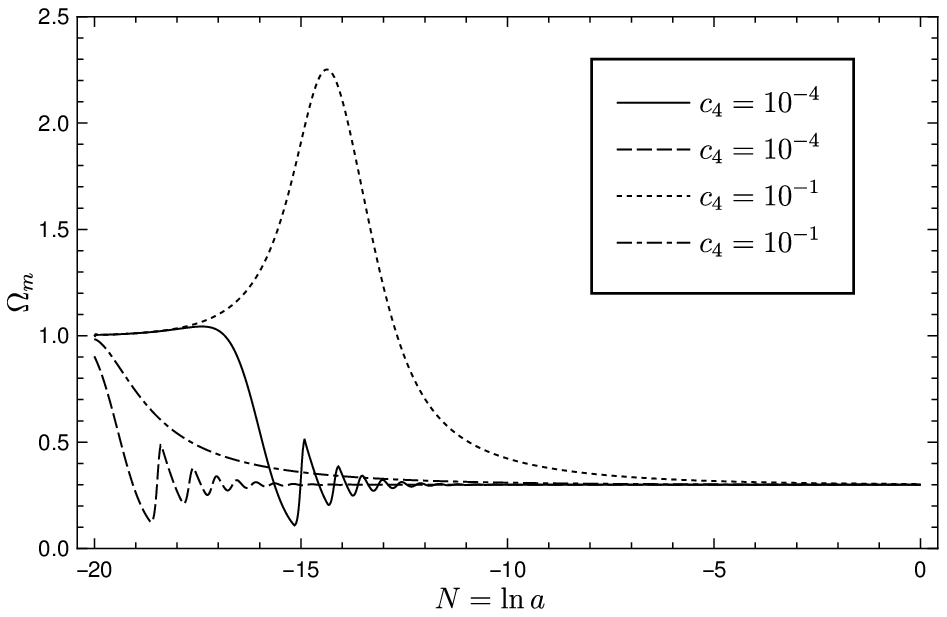}
\includegraphics[height=0.39\textwidth, width=0.49\textwidth,angle=0]{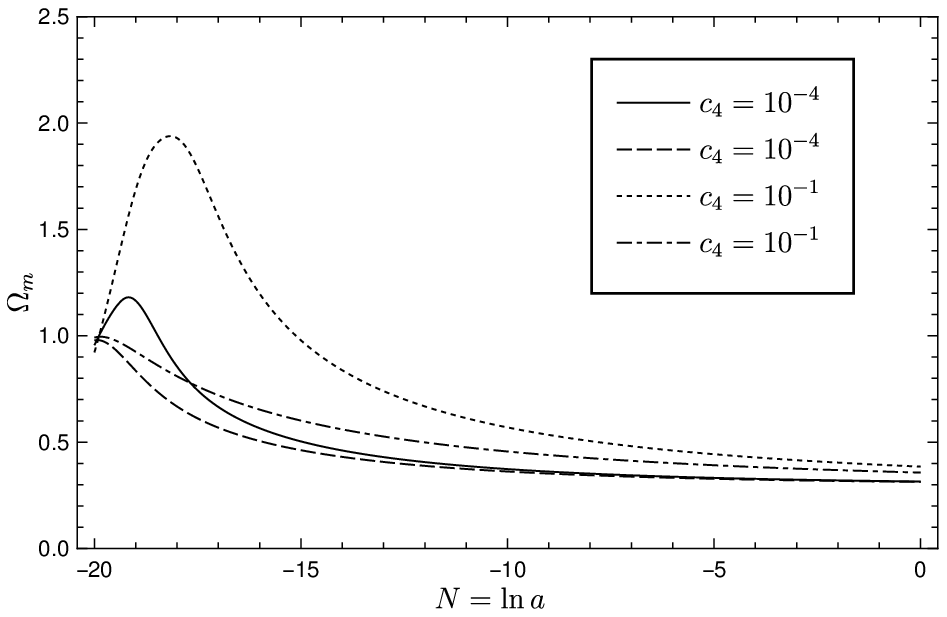}\\
\includegraphics[height=0.39\textwidth, width=0.49\textwidth,angle=0]{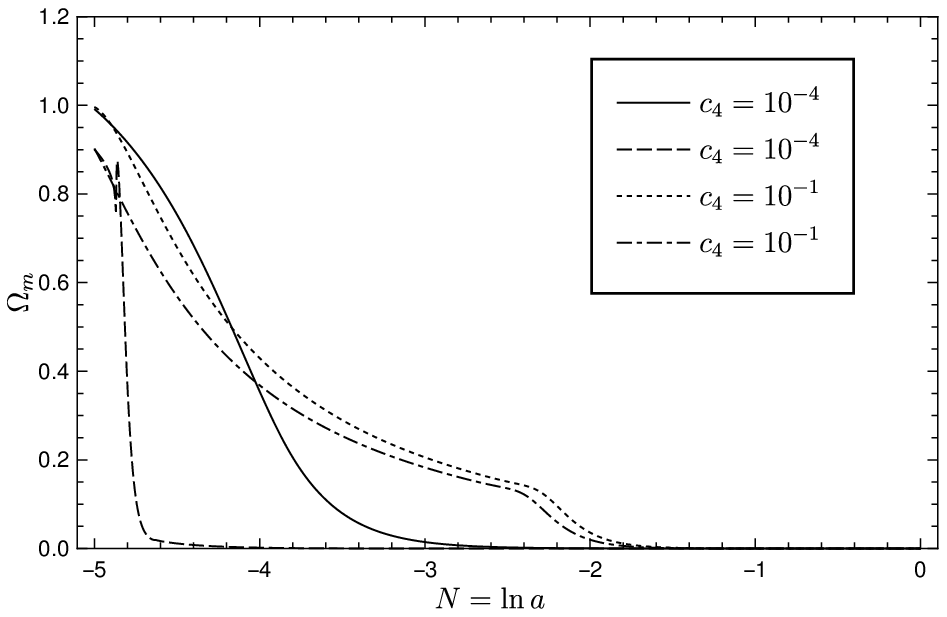}
\includegraphics[height=0.39\textwidth, width=0.49\textwidth,angle=0]{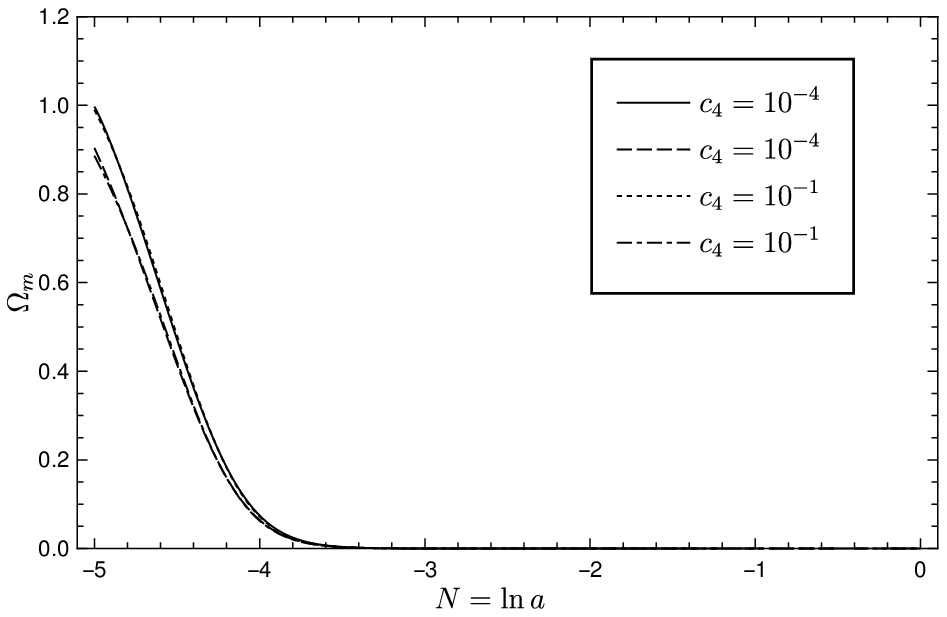}
\caption{\label{fig:2}
Plots of $\Omega_m$ as a function of $N$. The upper two panels represent the cases $x_r > 0$ during the matter domination, while the lower two panels
represent the cases $x_r < 0$ during the matter domination. The left two panels and the right two panels correspond to the model of $(n_2,n_4,n_6) = (0,-1,-1)$ and $(0,-1,1)$, respectively.
}
\end{figure}

Eq.~(\ref{fix:dh}) indicates that $\Omega_m = 0$ is a possible fixed point of the system.
To obtain Eq.~(\ref{fix:dh}), we set $h=x_\lm/2$ at the fixed point according to Eq.~(\ref{scalingSol}).
Nevertheless, the condition $h = x_\lm/2$ can be relaxed if $x_r,y_r$ and $z_r$ are not equal to unity at fixed point, where the condition $x_r=y_r=z_r=1$ defines the scaling fixed point.
From Eqs.~(\ref{xrp}) and (\ref{yrp}), we see that the fixed points exist when
\be
h = \frac{x_\lambda}{2} z_r = \frac{x_\lambda}{2} x_r\,,
\label{gpoint}  
\ee
where the expressions for $x_r$ and $z_r$ at the fixed point can be solved from Eqs.~(\ref{eiiless}), (\ref{e0less}) and (\ref{e1less}).
For the fixed point $\Omega_{m} = 0$, the expressions for $x_r$ and $z_r$are complicated and strongly depend on $n_2$, $n_4$ and $n_6$ because Eqs.~(\ref{eiiless}), (\ref{e0less}) and (\ref{e1less}) contain $x_r^{n_2}$, $x_r^{n_4}$ and $x_r^{n_6}$. 
However, we can substitute Eq.~(\ref{gpoint}) into Eq.~(\ref{weff}) to obtain
\be
w_\eff = w_\ef = -1 + \frac{x_\lambda x_{r\,b}}{3} \,,
\label{w:b}
\ee
where subscript ${}_b$ denotes evaluation at the field dominated point.
We note that for this fixed point there is no any requirement on $Q_\lambda$.
This follows from Eqs.~(\ref{drhop}) and (\ref{rmdot}) that the effect of the coupling $Q$ disappears when $\Omega_m = 0$.
According to this fixed point, the eigenvalues computed from the Jacobian matrix are given by
\be
E_b = \{\frac{x_\lm x_{r\,b}-6}{2}, x_\lm x_{r\,b}(Q_\lm + 1)-3\} \,.
\label{eb}
\ee
It follows from Eq.~(\ref{w:b}) that observational data require $x_\lambda x_{r\,b} < 1$ so that the first eigenvalue in Eq.~(\ref{eb}) is always negative.
We see that if $Q_\lm$ does not satisfy Eq.~(\ref{scalingcon}), the second eigenvalue in Eq.~(\ref{eb}) is negative when $Q_\lm < 3/(x_\lm x_{r\,b}) -1$ for positive $x_\lm x_{r\,b} $ and $Q_\lm > - 3/|x_\lm x_{r\,b}| -1$ for negative $x_\lm x_{r\,b}$.
These results are the same as in \cite{Scaling2006:1},
which implies that the modification of gravity theory has no effect on stability of the field dominated fixed point. 
In the case where $Q_\lm$ satisfies Eq.~(\ref{scalingcon}), one of the eigenvalues vanishes. 
In this case, the eigenvalues for the field dominated point are similar to those for the scaling fixed point which $c_4 \to \infty$. 
Since one of the eigenvalues vanishes, we cannot use the linear dynamical analysis to estimate the stability of the fixed point. 
However we will not go beyond the linear analysis in this work. 
For a given value of $x_\lm$ which could make the field dominated point stable, we can choose $n_2, n_4 , n_6$ and $c_4$ such that the scaling fixed point is also stable. 
The question is that the cosmic evolution will reach the scaling fixed point at late time in what situation.
Since it is difficult to perform analytical analysis for answering this question, we solve the autonomous equations numerically and plot the evolution of $\Omega_m$ in Fig.~(\ref{fig:2}) for some values of the model parameters.
According to Fig.~(\ref{fig:2}), the cosmic evolution will reach the scaling fixed point at late time if $x_r>0$ during the matter domination.
For $x_r<0$ during the matter domination, the cosmic evolution will evolve toward the field dominated point.
This result is consequences of a positive $x_\lambda$ of the scaling points given by Eq.~(\ref{dh2h2s}),
and the fact that the evolution of $x$ cannot cross $x = 0$.
This implies that although one of the eigenvalues vanishes, the field dominated point can be stable.
Since the scaling points we consider in the plots are stable, these points should be reached for wide ranges of initial conditions.
However, if $c_4$ is large enough and the initial condition for $y_r$ significantly differs from its value at the fixed point,
the value of $\Omega_m$ can be larger than unity before reaching the fixed point.
This implies that $\Omega_\eff$ can be negative,
so that the definitions in Eq.~(\ref{rpeff}) may cannot be interpreted as the energy density and pressure of dark component.
These cases are shown in Fig.~(\ref{fig:2}). 
In the top left panel of Fig.~(\ref{fig:2}), the initial values for $x_r$ and $y_r$ during the matter domination for the solid, long-dash, dash, and dash-long-dash lines are $(x_r,y_r) = (0.55,10^{-5}), (0.05,0.24), (0.1,10^{-8}), $ and $(0.79,0.7)$ respectively.
In the top right panel of Fig.~(\ref{fig:2}), the initial values for $x_r$ and $y_r$ during the matter domination for the solid, long-dash, dash, and dash-long-dash lines are $(x_r,y_r) = (0.4,0.2), (0.74,0.8), (0.18,0.01), $ and $(0.85,0.8)$ respectively.
For the cases where $y_r$ significantly differs from their values at the fixed point, the maximum value of $\Omega_m$ during the cosmic evolution increases when $c_4$ increases.
Since $c_4$ quantifies the deviation from the Einstein gravity,
this suggests that the deviation from the Einstein gravity should not be large to avoid unphysical value of $\Omega_m$ during the cosmic evolution.
Moreover, even though the initial values of $x_r$ and $y_r$ during the matter domination are in the same order of magnitude of the value at fixed point,
the cosmic evolution slower reaches the fixed point for positive initial $x_r$ compared with the negative initial value of $x_r$. 

\subsubsection*{(c) $y_r=0$ : $\phi$MDE point}

\begin{figure}
\includegraphics[height=0.7\textwidth, width=0.9\textwidth,angle=0]{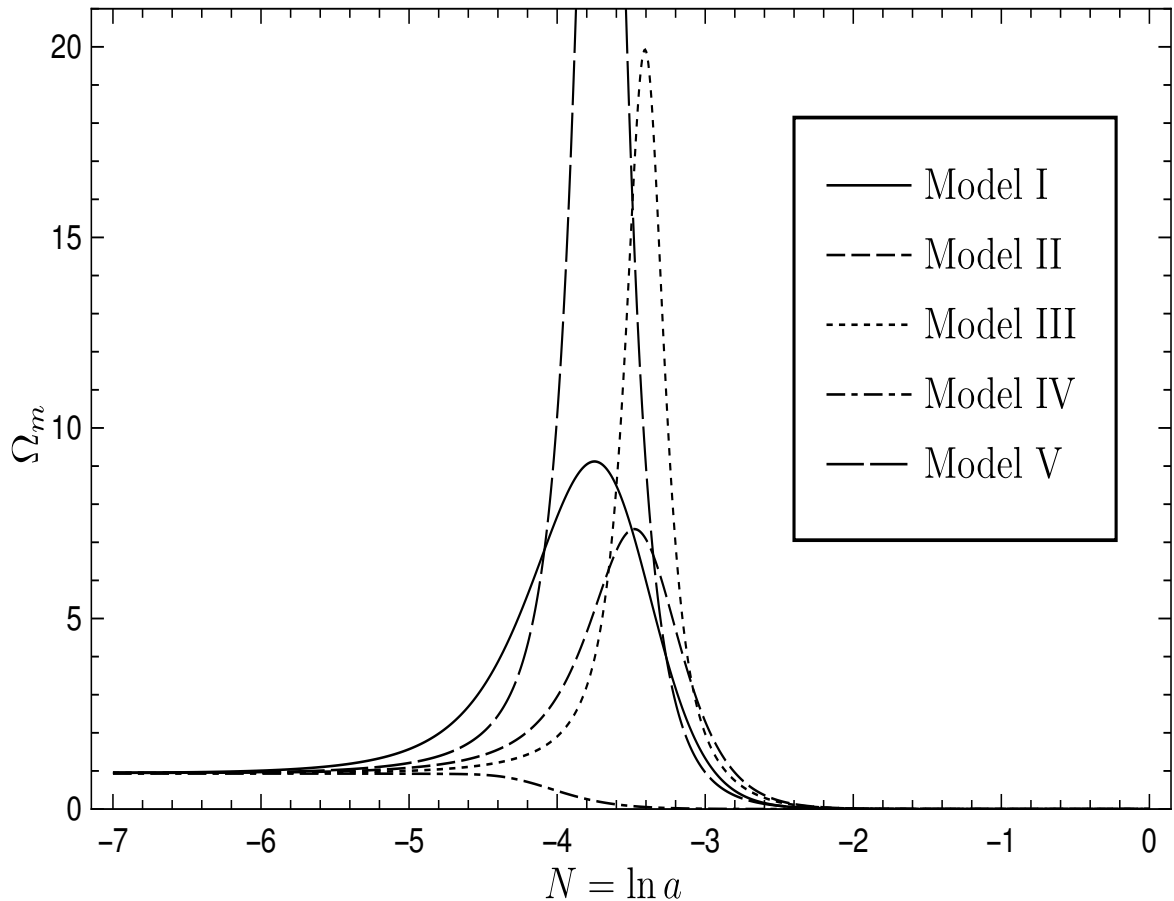}
\caption{\label{fig:3} Plots of $\Omega_{m}$ as a function of $N$ for models I--V given in the table.
}
\end{figure}

According to Eq.~(\ref{fix:dh}),
the other fixed point corresponds to $y_r =0 $.
It follows from Eq.~(\ref{yrp}) that $y_r' = 0$ when $y_r = 0$. 
If we consider Eq.~(\ref{xrp}) in addition, we see that $x_r'=0$ when $z_r = 2 h / x_\lambda$.
Here, $h$ for this fixed point is not necessarily equal to $ x_\lm/2$ because $x_\lm$ is evaluated at the scaling fixed point (fixed point a).
From the definitions of $G_2$ and $G_4$ in Eqs.~(\ref{G2}) and (\ref{G4}) as well as the definition of $y$ in Eq.~(\ref{dless}),
we see that the existence of the fixed point $y_r = 0$ requires $n_2 \leq 0, n_6 < 0$ and $n_4 < 0$.
Here, we demand that $n_2 \neq n_6$ and $n_4 \neq 0$. 
Inserting $z_r = 2 h / x_\lm$ and $\Omega_\ga = 0$ into Eqs.~(\ref{eiiless}), (\ref{e0less}) and (\ref{e1less}) and then taking the limit $y_r \to 0$,
we respectively obtain
\be
h|_c =  \frac{3 + c_2 x_{r\,c}^2}{2} \,, \quad  
\Omega_{m\,c} = 1 - \frac{c_2 x_{r\,c}^2}{3} \, \quad \mbox{and} \quad  x_{r\,c} = -\frac{Q_\lm x_\lm}{c_2} \,,
\label{fixc}
\ee
where the subscript ${}_c$ denotes evaluation at $\phi$MDE point.
Substituting the above $x_{r\,c}$ into the expression for $\Omega_{m\,c}$, we get
\be
\Omega_{m\,c} = 1 - \frac{Q_\lm^2 x_\lm^2}{3 c_2} \,.
\label{omc}
\ee
This equation shows that $c_2$ has to be positive otherwise $\Omega_{m\,c}$ is larger than unity.
The eigenvalues for this fixed point are
\be
E_c =\{-\frac{3}{2} + \frac{Q_\lm^2 x_\lm^2 }{2 c_2}  , 3 + \frac{Q_\lm(1+Q_\lm) x_\lm^2 }{c_2}  \} \,.
\ee
These eigenvalues coincide with those in \cite{Scaling2006:1}. 
The first eigenvalue can be written as $-3\Omega_{m\,c}/2$,
so that it is always negative. 
The second eigenvalue becomes positive when $Q_\lm > 0$ or $Q_\lm < -1$ for positive $c_2$. 
Since $x_\lm$ is evaluated at the scaling fixed point,
it follows from Eq.~(\ref{scalingcon}) that $Q_\lm < 1$ yields $x_\lm < 0$ corresponding to phantom expansion.
We now check how the evolution of the universe can move from this fixed point during matter domination to the scaling fixed point at late time.
Let us first consider $x_{r\,c}$ in Eq.~(\ref{fixc}). 
We can use Eq.~(\ref{scalingcon}) to write  $x_{r\,c} = (x_\lm -3)/c_2$. 
The scaling fixed point can lead to the acceleration of the universe if $x_\lm < 2$.
Hence, $x_{r\,c}$ is negative.
Since  $x_{r\,c}$ is the value of $x_r$ during matter domination in our consideration,
the universe will evolve toward the field dominated point rather than the scaling fixed point as presented in the previous section.
For illustration, we plot evolution of $\Omega_m$ in Fig.~(\ref{fig:3}).
For given values of $x_\lm$, $Q_\lm$, and  $\Omega_{m\,c}$,
the value of $c_2$ can be computed from Eq.~(\ref{omc}).
From the values of $x_\lm$, $Q_\lm$ and $c_2$,
we can compute $x_{r\,c}$ from Eq.~(\ref{fixc}) and compute $c_4$ from Eq.~(\ref{Eqc2}) by setting $\Omega_{m\,s} = 0.3$.
Finally, $c_6$ can be computed from Eq.~(\ref{Eqc6}).
The models used in the plots are shown in the table.
\begin{table*}[t]
\begin{center}\begin{tabular}{|c|c|c|c|c|c|}
    \hline \textbf{Model}	& $(n_2,n_4,n_6)$	& $Q_\lm$ 	& $x_{r\,c}$  	& $c_4 $ 	& $w_\ef$\tabularnewline
    \hline I			& (0,-1,-1)  		& -10		& -0.045 	& 7.7  		& 	-0.88	 \tabularnewline
    \hline II			& (0,-1,-1)  		&   2		& -0.075 	& 1.7 		& 	-1.28	 \tabularnewline
    \hline III 			& (0,-1,-1)  		&  2/3		& -0.125 	& 0.67  	& 	-1.44	 \tabularnewline
    \hline IV 			& (0,-1,-1)  		&  1/6		& -0.49 	& $5.6 \times 10^{-3}$	& -1.47 	\tabularnewline
    \hline V 			& (0,-1,-2)  		&   2		& -0.075  	& 4.0		& 	-1.17	 \tabularnewline
    \hline
\end{tabular}\end{center}
\caption{The models used in the plots. 
We set $\Omega_{m\,c}=0.95$ for Model I-IV and $\Omega_{m\,c}=0.93$ for Model V.
The column $w_\ef$ shows the value of $w_\ef$ at the field dominated point.
}
\end{table*}

From Fig.~(\ref{fig:3}), we see that $\Omega_{m}$ evolves toward the field dominated point for various values of $Q_\lm$ which correspond to various $w_\ef$ at late time.
In the plots, we initially set $y_r = 10^{-11}$ according to the $\phi$MDE point, so that
the value of $\Omega_{m}$ can be larger than unity before reaching the field dominated point. 
However, if $c_4$ is sufficiently small, e.g., $c_4 = 5.6 \times 10^{-3}$ for model IV, $\Omega_{m}$ can be less than unity through out the evolution of the universe.
By definition, $c_4$ quantifies how large of the deviation from the Einstein gravity.
The above results suggest that the deviation from the Einstein gravity should not be large to avoid the case $\Omega_m >1$ during the cosmic evolution.
From the analysis of the Vainshtein mechanism, the bound on the difference between the gravitational constant of the gravitational source and the gravitational coupling for GW gives \cite{Hirano2019:2}
\be
\left|\frac{X G_{4X}}{G_4} \right| < {\cal O}(10^{-3}) \,.
\ee
In terms of $c_4$, $|X G_{4X}| = |n_4 c_4|$ at the scaling fixed point.
Hence, the small $c_4$ seems to agree with the above bound.

\section{Stability of the linear perturbations}
\label{sec:4}

In this section,
we investigate stability of the linear perturbations in the theory considered in the previous sections around the FLRW background.
To describe the perturbations in the metric tensor,
we use the metric tensor in the ADM form,
\be
ds^2=- \n^2 dt^2 + h_{ij} (dx^i+\n^i dt)(dx^j+\n^jdt)\,,
\ee
 and quantify  the scalar perturbations in the unitary gauge by field variables $\de \n, \psi$ and $\zeta$ as
\ba
\n = 1 + \de \n\,, \quad 
\n^i=\de^{ij}\partial_j\psi\,,\quad 
h_{ij}=e^{2\zeta}\de_{ij}\,. 
\ea
The perturbed action for the DHOST theories up to the second order in perturbation is \cite{Langoi2017}
\be
\label{actzIa}
S^\quadac = \int d^3 x \, dt \, a^3   \frac{M^2 }{2} \bigg[ A_\tzeta  \dot{\tzeta}^2 
- B_\tzeta \frac{(\partial_i \tzeta)^2}{a^2} 
+ C_\tzeta \frac{(\partial_i \dot \tzeta)^2}{a^2}  \bigg] \; ,
\ee
where the propagating scalar degree of freedom is described by
\be
\tzeta \equiv \zeta - \beta_1 \delta \n\,.
\ee
The coefficients in the action (\ref{actzIa}) are given by
\begin{align}
A_\tzeta & = \frac{  1}{(1+\alphaB-\dot \bun /H)^2} \alpha \,,
 \label{AzI}\\
B_\tzeta & =  -2 (1+\alphaT)
+ \frac{2}{a M^2 }\frac{d}{dt}\bigg[\frac{a M^2 \big( 1+\alphaH+\bun(1+\alphaT)\big)}{H(1+\alphaB)-\dot \bun}\bigg]\,, 
\label{BzI}\\
C_\tzeta & = \frac{4(1+\alphaH) \bun + 2 (1+\alphaT) \bun^2 + \btrois}{(1+\alphaB-\dot \bun /H)^2} \,.
\end{align}
Here,
\be
\alpha \equiv \alphaK+ 6\alphaB^2 - \frac{6}{a^3 H^2 M^2} \frac{d}{dt} \left( a^3 H M^2 \alphaB \bun \right)\,.
\label{defal}
\ee
For the action in Eq.~(\ref{actGW2}),
the variables $M^2, \alphaK, \alphaB, \alphaH, \alphaL, \alphaT, \beta_1, \beta_2$ and $\beta_3$ defined in \cite{GLPV2015, Langoi2017} are
\ba
M^2 &=&  2 G_4\,, \quad 
\alphaT = 0\,, \quad
\alphaL = 0\,,
\\
\alphaB &=& 2 \beta_1
+ \frac{\dot{\phi } \left(6 G_{4 X} \ddot{\phi }+2 X G_{4 \phi X} + G_{4  \phi }\right)}{2 G_4 H}\,, 
\frac{4 H X G_{4 X}+\dot{\phi } \left(6 G_{4 X} \ddot{\phi }+2 X G_{4 \phi X}
+G_{4  \phi }\right)}{2 G_4 H}\,,
\\
\alphaH &=& -2 \beta_1\,, \quad
\beta_1 = \frac{X G_{4 X}}{G_4}\,,\quad
\beta_2 = - 3 \beta_1^2\,,\quad
\beta_3 = 6 \beta_1^2 -4 \beta_1\,,
\ea
and
\ba
\alphaK &=&
\frac{2 X \left(G_{2 X}+2 X G_{2 X X}\right)}{G_4 H^2}
-\frac{6 X \left(H \left(G_{4 X}+2 X G_{4 X X}\right)+\dot{\phi } \left(2 X G_{4 \phi  X X}+3 G_{4 \phi  X}\right)\right)}{G_4 H}
\nonumber\\
&& 
-\frac{36 \ddot\phi}{\dot\phi H}\beta_1
- \frac{6 \ddot\phi^2}{X H^2} \left( 4\beta_1^4 - \left(13 \beta_1 +10 X^2 \frac{G_{4 X X}}{G_4}\right) \beta_1^2
+ 15 \beta_1^2 + 2 X^2\left(13 \frac{G_{4 X X}}{G_4} + 2 X \frac{G_{4 X X X}}{G_4}\right) \beta_1 
\right.
\nonumber\\
&& \left.
+ 4 \left(X^2 \frac{G_{4 X X}}{G_4}\right)^2\right) \,.
\ea
According to Eqs.~(\ref{actzIa}) and (\ref{defal}),
the no ghost condition is
\be
\alpha > 0\,.
\label{alpha0}
\ee
We now add matter in our consideration
by supposing that the matter is described by K-essence \cite{Kase:2018iwp, Langoi2017} which the action is
\be
S_m = \int d^4 x \sqrt{-\tg} P(Y) \,, \qquad 
Y \equiv - \tg^{\mu \nu} \partial_\mu \sigma \partial_\nu \sigma\,,
\label{actsm}
\ee
where $\sigma$ is the scalar field and the metric $\tg_{\mu\nu}$ is related to the metric $g_{\mu\nu}$ in the action (\ref{actGW2}) by the conformal transformation,
\be
\label{disf_unit_I2_m}
\tg_{\mu \nu} = \e^{2 Q \phi} g_{\mu \nu}\,.
\ee
It can be shown that the action for the matter in the form of Eq.~(\ref{actsm}) can lead to Eq.~(\ref{rmdot}) (see e.g., \cite{vandeBruck:15, skd:18}).
Including the matter, the action (\ref{actzIa}) becomes \cite{Langoi2017, Crisostomi2019:2}
\be
S = \int d^3 x \,dt \, a^3   \frac{M^2 }{2} \bigg\{
\dot{\calv} {\cal K} \dot{\calv}^T
- \frac{1}{a^2}\partial_i\calv {\cal L} \partial^i\calv^T\,,
\bigg\}\,,
\label{s2phim}
\ee
where $\calv \equiv (\tzeta, \delta\sigma)$ and $\delta\sigma$ describes the perturbations in the field $\sigma$.
The matrix ${\cal K}$ and ${\cal L}$ are defined as
\be
{\cal K} \equiv \(\begin{array}{cc}\tilde{A}_\tzeta & B_m \\ B_m & A_m \end{array}\)\,,
\quad
{\cal L} \equiv \(\begin{array}{cc}B_\tzeta & C_m \\ C_m & A_m c_m^2\end{array}\)\,,
\label{matrices}
\ee
where $c_m^2 \equiv P_Y / (P_Y + 2 Y P_{YY})$ is the sound speed square of the matter perturbations,
a subscript ${}_Y$ denotes derivative with respect to $Y$, and
\begin{align}
A_m &\equiv  \frac{2 \e^{2 Q \phi} P_Y}{M^2 c_m^2}\,,
\\
\tilde{A}_\tzeta & =  A_\tzeta +\frac{{\cal M}}{\big[H(1+\alphaB)-\dot \bun \big]^2}\,,
\\
B_m & =  \frac{\e^{2 Q \phi} \rho_m (1 + w_m)}{\dsz M^2 c_m^2} \frac{1 - 3c_m^2\bun}
{H(1+\alphaB)-\dot \bun}\,,
\\
C_m & = \frac{\e^{2 Q \phi} \rho_m (1 + w_m)}{\dsz M^2} \frac{1+\alphaH+\bun(1+\alphaT)}
{H(1+\alphaB)-\dot \bun}\,,
\end{align}
The quantities $\dsz$ is the time derivative of the background field $\sigma_0$ and
\ba
{\cal M} &=& 3 \hat{\Omega}_m H^2 \[\(1+w_m\) \(6 \bun -9c_m^2 \bun^2\) +6 w_m \bun\] 
+ \frac{\e^{2 Q \phi} \rho_m (1 + w_m)}{c_m^2 M^2} \left[1 -3c_m^2 \bun\right]^2\,,
\\
\hat{\Omega}_m &=&  \frac{\e^{4 Q \phi}\left(2 Y P_Y - P\right)}{3 M^2 H^2}\,. 
\ea
The ghost and gradient instabilities can be avoided if the eigenvalues of the matrices ${\cal K}$ and ${\cal L}$ are positive.
From Eq.~(\ref{matrices}), the eigenvalues of these matrices can be written in the forms
\begin{align}
\lambda_{{\cal K}} &= \frac 12 \((\tilde{A}_\tzeta + A_m) \pm \sqrt{(\tilde{A}_\tzeta + A_m)^2 - 4(\tilde{A}_\tzeta A_m - B_m^2)}\)\,,
\\
\lambda_{{\cal L}} &= \frac 12 \((B_\tzeta + A_m c_m^2) \pm \sqrt{(B_\tzeta + A_m c_m^2)^2 - 4(B_\tzeta A_mc_m^2 - C_m^2)}\)\,.
\end{align}
We see that the model can be free from ghost instability if $\tilde{A}_\tzeta > 0$, $A_m > 0$ and $\tilde{A}_\tzeta A_m > B_m^2$.
Since $A_m$ is the coefficient of the kinetic term for the matter,
we expect that $A_m > 0$.
The coefficient $\tilde{A}_\tzeta > 0$ if Eq.~(\ref{alpha0}) is satisfied.
Using the limit $w_m \to 0$ and $c_e^2 \to 0$,
the condition $\tilde{A}_\tzeta A_m > B_m^2$ can be satisfied.
Hence, even though the matter is included in the consideration, the no ghost condition is still given by Eq.~(\ref{alpha0}).
Similar to the case of ghost instability,
the model can be free from gradient instability if $B_\tzeta > 0$, $A_m c_m^2 > 0$ and $B_\tzeta A_mc_m^2 > C_m^2$.
To check the latter condition,
we write
\be
B_\tzeta A_mc_m^2 - C_m^2
=
\frac{2 \e^{2 Q \phi} P_Y}{M^2 }\[B_\tzeta
-
\frac{3 \e^{2 Q \phi} \Omega_m }{M^2} \[\frac{1+\alphaH+\bun}{(1+\alphaB)-\bun'}\]^2 
\]
= \frac{2 \e^{2 Q \phi} P_Y}{M^2 }{\cal C}\,.
\ee
Hence, the gradient instability can be avoided if $B_\tzeta > 0$ and ${\cal C} > 0$.
We now estimate the factor $\e^{2 Q\phi}$ in the expression for ${\cal C}$.
It is difficult to estimate the value of $\phi$ from the results in the previous sections,
so  we estimate $Q$ from Eq.~(\ref{scalingcon}),
which gives $Q/\lambda \simeq  2.2$ if $x_\lm \simeq 0.92$.
Since the background dynamics descussed in the previous sections depend on $Q/\lambda$ rather than $Q$,
for a given value of $Q/\lambda$, we can choose $\lambda$ such that $Q$ is small or negative without altering the background dynamics.
Hence, for simplicity, we suppose that $\e^{2 Q \phi} \lesssim 1$.

We check the ghost and gradient instabilities at the scaling point which corresponds to the cosmic acceleration at late time.
We choose $x_\lm \simeq 0.92, \Omega_m = 0.3 $ and $c_4 = 0.1$.
We write $\alpha, B_\tzeta $ and ${\cal C}$ in terms of dimensionless variables and compute numerical values of these quantities.
We have found that there is no gradient instability and the ghost instability can be avoided if $n_6 > n_2$ for $ n_2 \geq 0 $ and for wide range of $n_4$. 
These conclusions do not depend on $c_4$.

 \section{Conclusions}
\label{sec:5}

In this work, we have studied the cosmic evolution based on the fixed points in the dynamical analysis of the DHOST theory which has the scaling solutions.
In addition to have the scaling solutions, the DHOST theory in our consideration satisfies the requirements that the propagation speed of GW is equal to the speed of light and the GW do not decay to dark energy perturbations.
We concentrate on the model parameters which the expression of $z_r$ is given by Eq.~(\ref{xr4sol1}).

We have found in our analysis that the scaling fixed point associated with the late time accelerating universe, is stable when $n_2$ and $n_6$ are not negative for $n_4 = -1$ and $-2$. 
The stability of this scaling fixed point also depend on the expansion rate of the universe at late time through the parameter $x_\lm$.
There are ranges of parameters in which the scaling fixed point and the field dominated point are simultaneously stable. 
The cosmic evolution will reach the scaling fixed point at late time if $x_r$ during the matter domination is positive.
If $x_r$ during the matter domination is negative, the cosmic evolution will reach the field dominated point.

When the scaling fixed point and the field dominated point are stable, these points can be reached at late time for wide ranges of $x_r$ and $y_r$ during the matter domination.
However, we have found that the density parameter of the matter can be larger than unity during the cosmic evolution if $c_4$ is large enough and the initial value of $y_r$ during the matter domination is significantly different from its value at those fixed points.
By definition, $c_4$ quantifies how large of the deviation from the Einstein gravity.
In our consideration, the allowed values of $c_4$ depend on the initial conditions for $x_r$ and $y_r$ during the matter domination. 

Even though the autonomous equations for the model considered here are different from coupled dark energy models presented in \cite{Scaling2006:1},
we have found that the eigenvalues for the field dominated and $\phi$MDE points in both model are the same.
In our numerical investigation, the universe can only evolve from the $\phi$MDE to the field dominated point.
We also have found that the eigenvalues for the scaling fixed point reduce to those for field dominated  point when $c_4$ is significantly large.
However, recall that the large $c_4$ can lead to unphysical values of $\Omega_m$ during the cosmic evolution.

We conclude that the fixed points for the DHOST theory studied in \cite{Scaling2019} are similar to those in the coupled dark energy model in \cite{Scaling2006:1}.
We have found that the eigenvalues for the field dominated and $\phi$MDE points in DHOST theory with scaling solutions are similar to those in coupled dark energy model even though  the theories of gravity in these models are different.
However, for DHOST theory,	 the expressions for the eigenvalues corresponding to the scaling point are complicated,
and consequently stability of the fixed point is evaluated numerically in this work.

We have also estimated the ghost and gradient instabilities in this theory.
We have found that this theory is free from the gradient instability,
while the ghost instability is absent when $n_6 > n_2$ for $n_2 \geq 0$ and for a wied range of $n_4$.

\subsection*{Acknowledgements}

WT was supported by Royal Thai Government Scholarship (Ministry of Higher Education, Science, Research and Innovation) for his Ph.D. study.
KK is supported by Fundamental Fund from National Science, Research and Innovation Fund under the grant ID R2565B030.

%%%%%%%%%%%%%%%%%%%%%%
\appendix
\section{Constraint equations in terms of dimensionless variables}

In terms of the dimensionless variables, we can write  Eq.~(\ref{eii}) as
\ba
  0 &=&  \frac{1}{2 c_4+v_r^{n_4}} \left[v_r^{-n_4} \left(v_r^{-n_2-n_6} \left(2 c_4+v_r^{n_4}\right) \left(2
    c_4 n_4 v_r^{n_2+n_6} \left(z_r x_{\lambda } \left(-\frac{\dot H}{H^2}+z_r
    x_{\lambda }-2\right) - x_{\lambda } z_r' + n_4 x_{\lambda }^2\right)
    \right. \right. \right.
     \nonumber\\
    &&
    \left. \left.\left.
    - x_r^2 v_r^{n_4} \left(c_6 v_r^{n_2}-c_2 v_r^{n_6}\right)\right)+\left(2
    c_4+v_r^{n_4}\right) \left(c_4 (4 \frac{\dot H}{H^2}+6)+\left(2
    \frac{\dot H}{H^2}+\Omega_\ga + 3\right) v_r^{n_4}\right)
    \right. \right. 
    \nonumber\\
    &&
    \left. \left.
    +c_4 n_4 z_r^2 x_{\lambda}^2 \left(c_4 \left(n_4-4\right)+2 \left(n_4-1\right) v_r^{n_4}\right)-4
    c_4 n_4^2 z_r x_{\lambda }^2 \left(2 c_4+v_r^{n_4}\right)
    \right. \right.
    \nonumber\\ 
    &&
    \left. \left.
    -c_4 n_4 x_{\lambda } \left(2 c_4+v_r^{n_4}\right) \left(z_r x_{\lambda}-4\right)\right)\right] \,,
    \label{eiiless} 
\ea
where $v_r \equiv y_r / x_r^2$.

Eq.~(\ref{e0:ori}) can be written in terms of the dimensionless variables as
\ba
 0 &=& \frac{1}{{\left(2 c_4+v_r^{n_4}\right)^2}}   \left[v_r^{-n_2-n_4-n_6} \left(-c_4^2 v_r^{n_4} \left(-4 c_6 \left(3 n_4-2
    n_6-1\right) x_r^2 v_r^{n_2}-4 c_2 \left(2 n_2-3 n_4+1\right) x_r^2 v_r^{n_6}
    \right. \right.\right.
    \nonumber\\
    &&
    \left. \left. \left.
    +3 v_r^{n_2+n_6} \left(4 n_4^3 x_r^2 x_{\lambda }^2+4 n_4 \left(2
    x_r x_{\lambda }-2 z_r x_{\lambda }+\Omega_\ga -2\right)+n_4^2 x_{\lambda }
    \left(x_r \left(8-2 z_r x_{\lambda }\right)
    \right. \right.\right.\right.\right.
    \nonumber\\
    &&
    \left. \left. \left.\left. \left.
    +z_r \left(z_r x_{\lambda}+4\right)\right)+12\right)\right)-2 c_4 v_r^{2 n_4} \left(-c_6 \left(3
    n_4-4 n_6-2\right) x_r^2 v_r^{n_2}-c_2 \left(4 n_2-3 n_4+2\right) x_r^2 v_r^{n_6}
    \right. \right.\right.
    \nonumber\\
    &&
    \left. \left. \left.
    +3 v_r^{n_2+n_6} \left(n_4 \left(x_r x_{\lambda }-z_r x_{\lambda}+\Omega_\ga-1\right)+3\right)\right)-6 c_4^3 v_r^{n_2+n_6} \left(n_4^3
    x_{\lambda }^2 \left(-4 x_r z_r+4 x_r^2-z_r^2\right)
    \right. \right.\right.
    \nonumber\\
    &&
    \left. \left. \left.
    +n_4^2 x_{\lambda } \left(x_r \left(8-2 z_r x_{\lambda }\right)+z_r \left(z_r x_{\lambda }+4\right)\right)+4 n_4 \left(x_{\lambda }
    \left(x_r-z_r\right)-1\right)+4\right)
    \right. \right.
    \nonumber\\
    &&
    \left. \left.     
    +v_r^{3 n_4} \left(-c_6 \left(2
    n_6+1\right) x_r^2 v_r^{n_2}+c_2 \left(2 n_2+1\right) x_r^2 v_r^{n_6}-3
    v_r^{n_2+n_6}\right)\right)\right] +3
    \left(\Omega _m+\Omega_\ga \right) \,.
 \label{e0less}
\ea 
This equation can be used to express $\Omega_m$ in terms of the other dimensionless variables.
Eq.~(\ref{eti}) can be written in terms of the dimensionless variables as
\ba
0 &=&  v_r^{-n_2-4 n_4-n_6} \left(c_4^3 \left(-6 n_4^4 \left(8 x_r^3-18 z_r
    x_r^2+9 z_r^2 x_r+z_r^3\right) x_{\lambda }^3 v_r^{n_2+n_6}
    \right. \right.
    \nonumber\\
    &&
    \left. \left.  
    +3 n_4^3 x_{\lambda }^2 \left(8 x_{\lambda } x_r^3-96 x_r^2-3 z_r \left(z_r
    x_{\lambda }-24\right) x_r+z_r^2 \left(z_r x_{\lambda }+12\right)\right) v_r^{n_2+n_6}
    \right. \right.
    \nonumber\\
    &&
    \left. \left.  
    +12 n_4 \left(c_6 x_r^2 \left(\left(2 n_6-1\right) x_r x_{\lambda }-\left(2 n_6 z_r+z_r\right) x_{\lambda }+8\right)
    v_r^{n_2}-c_2 x_r^2 \left(\left(2 n_2-1\right) x_r x_{\lambda }
    \right. \right.\right. \right.
    \nonumber\\
    &&
    \left.\left.\left.\left.   
    -\left(2 n_2 z_r+z_r\right) x_{\lambda }+8\right) v_r^{n_6}+\left(x_r-z_r\right)
    \left(\Omega_\ga -3\right) x_{\lambda } v_r^{n_2+n_6}\right)-12 n_4^2
    x_{\lambda } \left(c_6 x_r^2 \left(2 x_r+z_r\right) v_r^{n_2}
    \right. \right.\right. 
    \nonumber\\
    &&
    \left.\left.\left.
    -c_2 x_r^2 \left(2 x_r+z_r\right) v_r^{n_6}+\left(x_{\lambda } \left(z_r x_{\lambda}
    -4\right) x_r^2-2 \left(\Omega_\ga + 2 z_r x_{\lambda }-7\right) x_r
    \right. \right.\right.\right.  
    \nonumber\\
    &&
    \left.\left.\left.\left.
    +z_r \left(-\Omega_\ga + 2 z_r x_{\lambda }+4\right)\right) v_r^{n_2+n_6}\right)+8
    x_r \left(3 v_r^{n_2+n_6} \left(\Omega_m + \Omega_\ga \right) Q_{\lambda } x_{\lambda }
    \right. \right.\right.
    \nonumber\\
    &&
    \left.\left.\left.
    -x_r \left(c_6 v_r^{n_2} \left(2 \left(x_r-z_r\right)
    x_{\lambda } n_6^2+\left(x_r x_{\lambda }-3 z_r x_{\lambda }+6\right)
    n_6-z_r x_{\lambda }+6\right)-c_2 v_r^{n_6} \left(2 \left(x_r-z_r\right) x_{\lambda } n_2^2
    \right. \right.\right.\right.\right.   
    \nonumber\\
    &&
    \left.\left.\left.\left.\left.
    +\left(x_r x_{\lambda }-3 z_r x_{\lambda }+6\right) n_2-z_r x_{\lambda }+6\right)\right)\right)\right) v_r^{n_4}-3 c_4^2
    \left(8 n_4^4 x_r^2 \left(x_r-z_r\right) x_{\lambda }^3 v_r^{n_2+n_6}
    \right. \right.
    \nonumber\\
    &&
    \left.\left.
    -2 n_4^3 x_{\lambda }^2 \left(x_{\lambda } x_r^3+\left(z_r x_{\lambda
    }-16\right) x_r^2+8 z_r x_r+2 z_r^2\right) v_r^{n_2+n_6}+n_4 \left(-4 c_6
    x_r^2 \left(\left(2 n_6-1\right) x_r x_{\lambda }
    \right. \right.\right.\right.   
    \nonumber\\
    &&
    \left.\left.\left.\left.
    -\left(2 n_6 z_r+z_r\right) x_{\lambda }+8\right) v_r^{n_2}+4 c_2 x_r^2 \left(\left(2
    n_2-1\right) x_r x_{\lambda }-\left(2 n_2 z_r+z_r\right) x_{\lambda}+8\right) v_r^{n_6}
    \right. \right.\right.   
    \nonumber\\
    &&
    \left.\left.\left.
    -2 \left(x_r-z_r\right) \left(2 \Omega_\ga -3\right)
    x_{\lambda } v_r^{n_2+n_6}\right)+n_4^2 x_{\lambda } \left(6 c_6 x_r^2 z_r
    v_r^{n_2}-6 c_2 x_r^2 z_r v_r^{n_6}+\left(x_{\lambda } \left(z_r x_{\lambda }-4\right) x_r^2
    \right. \right.\right.\right.
    \nonumber\\
    &&
    \left.\left.\left.\left.
    +\left(8-4 z_r x_{\lambda }\right) x_r+2 z_r
    \left(-3 \Omega_\ga + z_r x_{\lambda }+5\right)\right) v_r^{n_2+n_6}\right)-4
    x_r \left(3 v_r^{n_2+n_6} \left(\Omega_m+\Omega_\ga \right) Q_{\lambda } x_{\lambda }
    \right. \right.\right.   
    \nonumber\\
    &&
    \left.\left.\left.
    -x_r \left(c_6 v_r^{n_2} \left(2 \left(x_r-z_r\right)
    x_{\lambda } n_6^2+\left(x_r x_{\lambda }-3 z_r x_{\lambda }+6\right)
    n_6-z_r x_{\lambda }+6\right)-c_2 v_r^{n_6} \left(2 \left(x_r-z_r\right)
    x_{\lambda } n_2^2
    \right. \right.\right.\right.\right.   
    \nonumber\\
    &&
    \left.\left.\left.\left.\left.
    +\left(x_r x_{\lambda }-3 z_r x_{\lambda }+6\right)
    n_2-z_r x_{\lambda }+6\right)\right)\right)\right) v_r^{2 n_4}
    \right. 
    \nonumber\\
    &&
    \left.
    -3 c_4 \left(2 \left(x_r-z_r\right) \left(-c_6 x_r^2 v_r^{n_2}+c_2 x_r^2
    v_r^{n_6}+\left(\Omega_\ga-1\right) v_r^{n_2+n_6}\right) x_{\lambda }
    n_4^2+\left(-c_6 x_r^2 \left(\left(2 n_6-1\right) x_r x_{\lambda }
    \right. \right.\right.\right.  
    \nonumber\\
    &&
    \left.\left.\left.\left.
    -\left(2 n_6 z_r+z_r\right) x_{\lambda }+8\right) v_r^{n_2}+c_2 x_r^2 \left(\left(2
    n_2-1\right) x_r x_{\lambda }-\left(2 n_2 z_r+z_r\right) x_{\lambda }+8\right) v_r^{n_6}
    \right. \right.\right.   
    \nonumber\\
    &&
    \left.\left.\left.
    -\left(x_r-z_r\right) \left(\Omega_\ga -1\right)
    x_{\lambda } v_r^{n_2+n_6}\right) n_4-2 x_r \left(3 v_r^{n_2+n_6}
    \left(\Omega_m+\Omega_\ga \right) Q_{\lambda } x_{\lambda }
    \right.\right.\right.  
    \nonumber\\
    &&
    \left.\left.\left.
    -x_r \left(c_6 v_r^{n_2} \left(2 \left(x_r-z_r\right) x_{\lambda } n_6^2+\left(x_r
    x_{\lambda }-3 z_r x_{\lambda }+6\right) n_6-z_r x_{\lambda }+6\right)-c_2
    v_r^{n_6} \left(2 \left(x_r-z_r\right) x_{\lambda } n_2^2
    \right. \right.\right.\right.\right.   
    \nonumber\\
    &&
    \left.\left.\left.\left.\left.
    +\left(x_r x_{\lambda }-3 z_r x_{\lambda }+6\right) n_2-z_r x_{\lambda
    }+6\right)\right)\right)\right) v_r^{3 n_4}+x_r \left(3 v_r^{n_2+n_6}
    \left(\Omega_m + \Omega_\ga \right) Q_{\lambda } x_{\lambda }
    \right. \right. 
    \nonumber\\
    &&
    \left.\left.
    -x_r \left(c_6 v_r^{n_2} \left(2 \left(x_r-z_r\right) x_{\lambda } n_6^2+\left(x_r
    x_{\lambda }-3 z_r x_{\lambda }+6\right) n_6-z_r x_{\lambda }+6\right)-c_2
    v_r^{n_6} \left(2 \left(x_r-z_r\right) x_{\lambda } n_2^2
    \right. \right.\right.\right.
    \nonumber\\
    &&
    \left.\left.\left.\left.
    +\left(x_r x_{\lambda }-3 z_r x_{\lambda }+6\right) n_2-z_r x_{\lambda
    }+6\right)\right)\right) v_r^{4 n_4}
    \right.   
    \nonumber\\
    &&
    \left.
    +6 c_4^4 n_4 x_{\lambda } \left(4
    n_4^2 x_{\lambda }^2 x_r^3+2 n_4 x_{\lambda } \left(2 z_r x_{\lambda }
    n_4^2-2 \left(z_r x_{\lambda }+8\right) n_4-z_r x_{\lambda }+4\right) x_r^2
    \right. \right.   
    \nonumber\\
    &&
    \left.\left.
    -\left(12 z_r^2 x_{\lambda }^2 n_4^3+z_r x_{\lambda } \left(3 z_r
    x_{\lambda }-40\right) n_4^2-8 \left(z_r x_{\lambda }-4\right)
    n_4+4\right) x_r
    \right. \right.   
    \nonumber\\
    &&
    \left.\left.
    -\left(n_4-1\right) z_r \left(n_4 z_r x_{\lambda}-2\right){}^2\right) v_r^{n_2+n_6}\right)  \,.
\label{e1less} 
\ea
To compute the equation for $z_r$, we substitute $\Omega_m$ solved from Eq.~(\ref{e0less}) into the above equation. the resulting equation can be written in the form
\be
b_3 z_r^3 + b_2 z_r^2 + b1 z_r + b_0 = 0 \,,
\label{xr4eq}
\ee
where $ b_0 ,b_1, b_2$ and $b_3$ are complicated funtions of the dimensionless variables of $\Omega_\ga, x_r, y_r $ and $x_\lm$.
Using Eq.~(\ref{xr4eq}), we can compute the expression for $z_r$ in the form
\be
 z_{r\,1} =  -\frac{\sqrt[3]{2} \left(3 b_1 b_3-b_2^2\right)}{3 b_3 \sqrt[3]{\Delta}}+\frac{\sqrt[3]{\Delta}}{3 \sqrt[3]{2} b_3}-\frac{b_2}{3 b_3}\,,
\label{xr4sol1}
\ee
\be
 z_{r\,2} =  \frac{\left(1+i \sqrt{3}\right) \left(3 b_1 b_3-b_2^2\right)}{3 (2^{2/3} b_3 \sqrt[3]{\Delta})}-\frac{\left(1-i \sqrt{3}\right) \sqrt[3]{\Delta}}{6 \sqrt[3]{2} b_3}-\frac{b_2}{3 b_3} \,,
\label{xr4sol2}
\ee
\be
 z_{r\,3} =\frac{\left(1-i \sqrt{3}\right) \left(3 b_1 b_3-b_2^2\right)}{3 (2^{2/3} b_3 \sqrt[3]{\Delta})}-\frac{\left(1+i \sqrt{3}\right)\sqrt[3]{\Delta}}{6 \sqrt[3]{2} b_3}-\frac{b_2}{3 b_3}\,,
\label{xr4sol3}
\ee
where $\Delta=  -2 b_2^3+9 b_1 b_3 b_2-27 b_0 b_3^2+\sqrt{4 \left(3 b_1 b_3-b_2^2\right){}^3+\left(-2 b_2^3+9 b_1 b_3 b_2-27 b_0 b_3^2\right)^2}$.
The physically relevant solution is selected from the above solutions by the requirement that $z_r$ becomes unity when $x_r=y_r=1, \Omega_\ga = 0$ and $c_2$ as well as $c_6$ are given by Eqs.~(\ref{Eqc2}) and (\ref{Eqc6}).

\end{document}